# Generic failure mechanisms in adhesive bonds

## P. Hass*, F.K. Wittel, P. Niemz

Institute for Building Materials, ETH Zurich, Schafmattstrasse 6, CH-8093 Zurich

*phass@ethz.ch

## Abstract

The failure of adhesive bondlines has been studied at the microscopic level via tensile tests. Stable crack propagation could be generated by means of samples with improved geometry, which made in-situ observations possible. The interaction of cracks with adhesive bondlines under various angles to the crack propagation was the focus of this study as well as the respective loading situations for the adhesives UF, PUR, and PVAc, which have distinctly different mechanical behaviors. It is shown how adhesive properties influence the occurrence of certain failure mechanisms and determine their appearance and order of magnitude. With the observed failure mechanisms, it becomes possible to predict the propagation path of a crack through the specimen.

## Introduction

Cellular failure mechanisms in bulk wood have been the subject of many investigations in the past for various wood species as well as failure in all anatomical directions, for single as well as mixed-mode loading (Borgin 1971; Bodner et al. 1997a,b; Thuvander and Berglund 2000; Tschegg et al. 2001; Dill-Langer et al. 2002; Reiterer and Sinn 2002; Conrad et al. 2003; Koponen and Tukiainen 2006; Keunecke et al. 2007; Vazi and Stanzl-Tschegg 2007; Stanzl-Tschegg and Navi 2009; Oliveira et al. 2009). While studies on solid wood have been carried out on all scales of length, down to tests on single fibers (Eder et al. 2008), knowledge on failure mechanisms of adhesive bonds is mainly based on large sized samples such as the double cantilever beam (Singh et al. 2010; Dourado et al. 2010; Nicoli et al. 2012). Estimates on microscopic failure mechanisms has mainly resulted from indirect observations like fracture surface investigations (River et al. 1994; Simon and Valentin 2000; Simon and Valentin 2003), video image correlation of the sample surface (Niemz et al. 2007), or accoustic emissions during failure (Suzuki and Schniewind 1987).

These surveys primarily focused on fracture mechanical properties, ignoring underlying generic failure mechanism of adhesive bonds. However, it is commonly known that failure is initiated on a small scale by micro defects that interact and join to form cracks that grow and become relevant on a larger scale. Depending on the adhesive type, moisture induced stresses resulting from hindered swelling and shrinkage, as well as cracks that develop during the curing of an adhesive, induce defects into the bonding (River 2003; Frihart 2009). However, failure mechanisms and crack evolution in adhesive bonds have not yet been studied for a constant climate. Investigations at the microscopic or mesoscopic scale are therefore essential to develop an understanding of the behavior and failure of wood adhesive connections that are of fundamental importance for modern wood constructions under various loading situations, made with different adhesive systems.

In the present study, microscopic failure mechanisms in adhesive bonds, made of systems with differing elasticity and curing reactions, will be studied. To this aim, the crack propagation and crack-bondline interaction will be observed in-situ under mode I loading. In addition to the effect of adhesive properties, the influence of the bondline orientation on the crack initiation direction will be observed.

## Material and methods

### Sample material and preparation

For the current investigation, a necked sample shape, as used by Dill-Langer et al. (2002), was preferred over other micro-test set-ups known for solid wood under mode I (Keunecke et al. 2007; Frühmann et al. 2003). To increase the crack growth stability, the oak wood supports proposed by Dill-Langer et al. (2002) for the load transfer into the test section were replaced by aluminum supports bonded by a PUR-adhesive (Figure 1c).

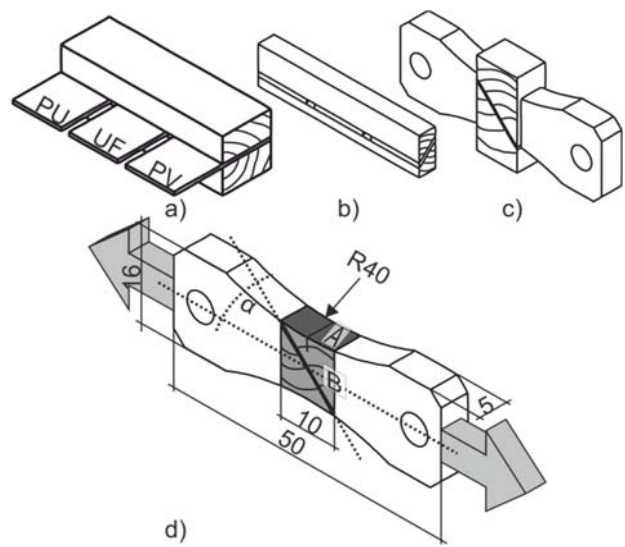

**Figure 1:** Sample preparation procedure: a) bonding of beams with three different adhesives; b) planning and aligning the BL; c) joining of wood and aluminum supports; d) finished specimen with dimensions [mm]; arrows indicate the load direction, α indicates the angle between load direction and BL, here 45°; darker surfaces represent different preparation techniques: A: belt-grinder, B: microtome.

Specific characteristics as well as the recommended processing conditions of the adhesive system investigated on spruce (*Picea abies* (L.) Karst) are summarized in Table 1.

**Table 1:** Adhesive systems, properties, and processing conditions.

| Adhe-sive | Description and application[1] | Amount [g/m$^2$][2] | Open time [min][1] | Press time [min][1] | Pressure [MPa][1] | Solid content [%][1] | MOE [MPa][3] |
|---|---|---|---|---|---|---|---|
| PUR | 1K-PUR for structural wood products | 200 | 40 | 100 | **0.7** | 100 | 1190 |
| PVAc | Adhesive dispersion for universal application in timber industry (D3) | 200 | 8 | 10 | Minimum 0.25 | 50-52 | 530 |
| UF | Cold-setting adhesive powder (EN 12765 C3) containing hardener | 200 | 20 | **480** | Minimum 0.25 | 60 | 3000 |

[1] Manufacturers' declarations for 20°C; [2] Within range of manufacturers' recommendations; [3] Obtained from own compression tests on adhesive cubes for UF and tensile tests on adhesive films for PUR and PVAc, respectively.

Wood beams with a length of 500 mm were bonded by applying the three adhesives, each along a third of the total length along the longitudinal beam direction (Figure 1a). The variation of the wood properties in this direction is low and the two wooden pieces could be bonded simultaneously, which ensured constant press parameters for all samples. The applied pressure and the press time were determined for the respective adhesive with the highest requirements (bold print in Table 1). Each pair of adherends was derived from one beam, which had been divided in half, before both parts were bonded again. After curing and acclimatization at 20°C and 65% RH, the bonded beams were planed to cross-section dimensions of 10 x 20 mm$^2$ (Figure 1b). Crack propagation through wood is most unstable under a TR configuration (i.e., load in tangential, T, and crack growth in radial, R, direction; Gross and Seelig 2007; Bodig and Jayne 1982). Therefore, this worst case situation was chosen for this investigation to determine the influence of the bondline (BL). The angle between load and BL was varied (0°, 45°, 60°, 75°, 90°). From each beam, five sections of 7 mm length were taken from each adhesive region. The overlapping cross-sections of the samples were removed by a sledge microtome GSL 1 (WSL Switzerland), to allow for microscopic in-situ observation (Figure 1d). In total, 225 (3 adhesives x 3 beams x 5 angles x 5 repetitions) samples with adhesive bondlines were tested. A radial crack initiation notch was introduced on one sample side either from pith to bark (in-radial) or in the opposite direction (contra-radial). As reference, 15 unbonded specimens from three different beams were tested. The samples were loaded in a Deben Microtest microstage via alignment pins, which only allowed for rotation around their axis as the crack propagated. This way, the maximum load was always at the crack tip and momentum influences were minimized. In situ observations were made with a stereo-microscope, at a frequency of 5 Hz at a loading rate of 0.1 mm min$^{-1}$. Note that imaging was triggered at a load of 10 N and that measured force-displacement curves were used for synchronization with the images.

**Analysis**

The acquired images were evaluated first qualitatively focusing on the different adhesives. For each beam, a direct comparison between adhesives could be achieved for in-radial (IR) and contra-radial (CR) crack growth, as well as between the adhesives themselves. This information was then used to find differences between the BL-load angles. As such samples are too small to reliably measure fracture mechanical properties, the load-displacement data was considered only as an indicator.

# Results and discussion

**Solid wood (SW)**

In Figure 2$_I$, the typical failure mechanisms for mode I loading in the T-direction are given along with the respective load-displacement curve (Figure 2 b).

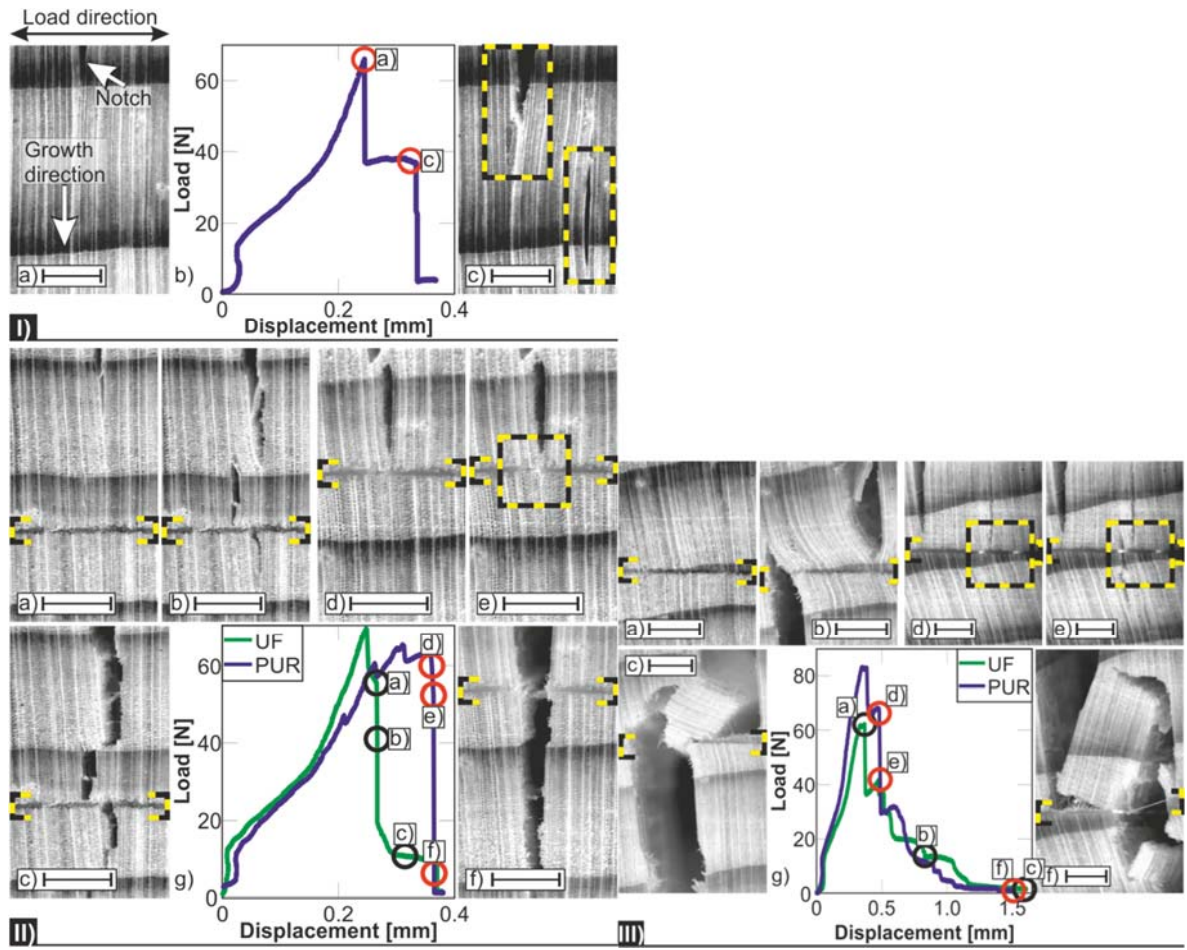

**Figure 2:** Crack propagation (CP) through spruce wood. All scales correspond to 1 mm. The plots show the load displacement diagrams. I) SW with IR CP. Maximum loading (a), and post-peak behavior (c). II) BL at 0° with CR CP through UF (a-c) and PUR (d-f) without deflection at BL; box in e) shows stress-whitening in adhesive layer; III) BL at 0° with CR CP through UF (a-d) and IR CP through PUR (e-g) with deflection at BL; boxes highlight stress whitening and adhesive fingers.

At the beginning of all load-displacement curves, adjustment processes of the microstage dominate. Note that all images of different failure situations are aligned in the same way concerning the load direction. The cracks advanced rather straight through the samples. As visible in the load-displacement curve (Figure 2$_I$ b), samples failed relatively abruptly, with a distinct pre-cracking (Figure 2$_I$ c) in the latewood (LW) zone, often with a parallel offset to the main crack. Intercellular fracture could be observed in the LW, while intracellular fracture dominated in earlywood (EW). This supports the known failure mechanisms for solid spruce wood under the load situation applied in this paper (Thuvander and Berglund 2000; Dill-Langer et al. 2002; Conrad et al. 2003). Since no influence of the crack initiation direction was observed for any of the bonded samples, this factor is disregarded in further discussion.

**Bonded samples**

Three distinct orientations were found to capture typical failure situations: bond lines (BL) parallel (0°) and perpendicular (90°) to the loading direction, as well as angles in between (45°-75°). In the following, samples from identical beams are juxtaposed for each of these cases to highlight the differences between the adhesive systems. Existing BLs in the images are highlighted by brackets. Since PUR and PVAc showed quite similar behavior regarding the crack propagation, often only one representative sample is presented. Additionally, the influence of pre-damages at BLs and the observations of adhesive layer delamination are discussed before the quantitative subsumption of the results.

***Load direction and BL at 0°:*** The BL has different properties than the adjacent wood, and hence can be compared to an additional growth ring border. As this layer was orientated in the T-direction, the final failure pattern showed features similar to SW: the crack could cross the BL in a more or less straight line (Figure 2$_{II}$) or it could be deflected at the BL, leading to roll-shear failure along the BL or along a growth ring border, if it is adjacent to the BL (Figure 2$_{III}$). This roll-shear failure is also typical for SW, if shifted pre-cracks form ahead of the main crack, leading to failure in the EW zone along growth ring borders. For the three adhesives, crack propagation and crack-BL interactions differed, as described in the following.

In UF bonded samples, the brittle BL actually acted as an additional latewood zone, where pre-cracks originated, leading to preferred paths for the main crack (Figure 2$_{II}$ a-c). Former studies (Hass et al. 2010) revealed a distinct crack pattern in the adhesive layer due to the restrained shrinkage in UF BLs during hardening. It therefore can be assumed that the relevant pre-crack in the BL emanated from the curing of the adhesive.

In PUR or PVAc BLs, pre-cracks were not detectable, since these systems are softer (see Table 1). In most cases, the crack stopped at the BL before penetrating it (Figure 2$_{II}$ d-f). In many cases the BLs even stayed intact, while the crack continued below it, either directly or via a pre-crack, which appeared in a latewood zone across the BL that grew upwards into the BL and downwards through the sample (Figure 2$_{III}$ d-f). Further in the failure progress, the adhesive layer dissipated energy, which became visible via stress-whitening, a common change in translucency of polymeric materials (Figures 2$_{II}$ e, 2$_{III}$ d-e).

If a pre-crack tangentially shifted with respect to the main crack at the BL, the coalescing of the cracks resulted in a roll-shear failure pattern. Here again the adhesives showed different reactions. In UF, the same behavior as for SW was observed: the EW next to the growth ring border or in the BL was sheared off with some fiber bridging (Figure 2$_{III}$ a-c). In PUR, the failure path followed the BL, while distinct adhesive fingers formed (Figure 2$_{III}$ d-f). However, the quantitative differences are small (Figure 2$_{III}$ g).

***Load direction and BL at 45°-75°:*** The influence of the BL grew with increasing angle between BL and load direction or decreasing angle between crack growth direction and BL. Consequently, the probability of crack deflection at the BL increased for higher angles. Distinct differences between the adhesives could be observed here, as a critical angle seemed to exist at which a crack deflection at the BL became more probable or, in other terms, energetically more favorable than BL penetration. For PUR, most cracks had already deflected at the BL for load-BL angles of 45°, while for PVAc, the number of deflections increased for angles above 60° and only at 75° did at least half of the samples show a deflection for UF (Table 2).

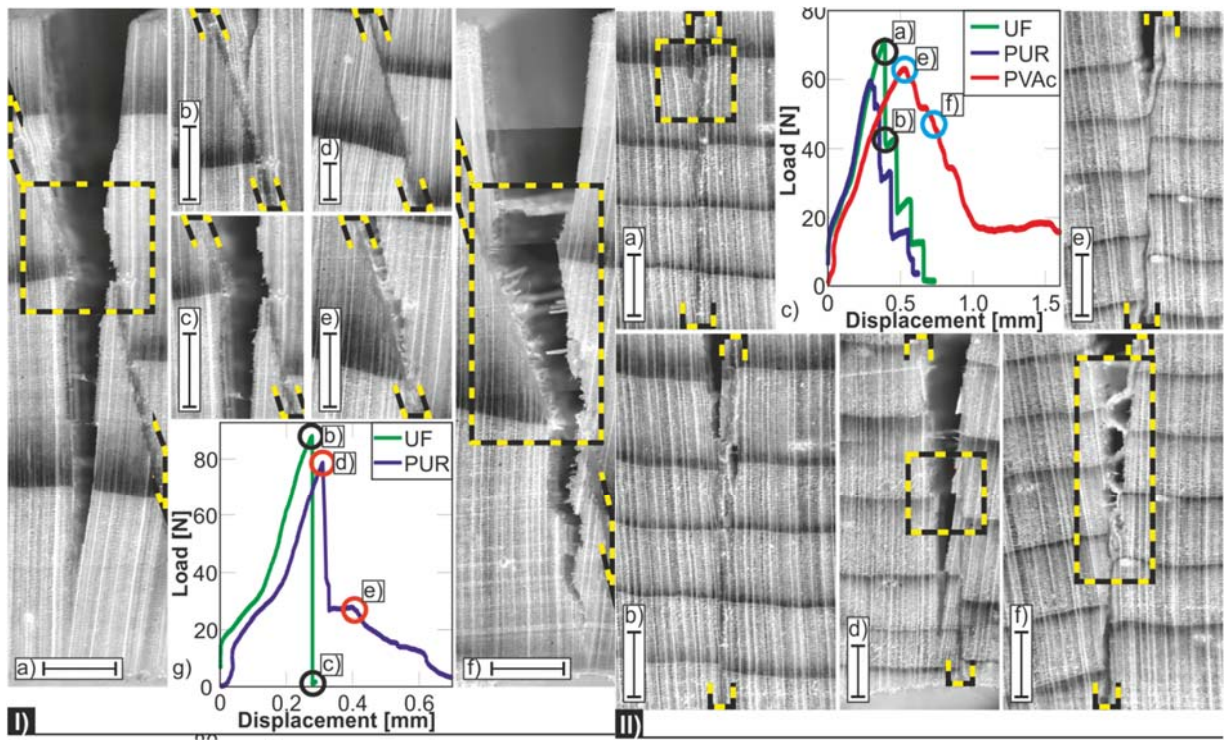

**Figure 3: I)** BL at 75° with IR CP through UF (a-c) and PUR (d-g); a) and f) show final stages of failure. Boxes highlight the deflected crack path until crack crosses BL at growth ring border. **II)** BL at 90°. a-b) IR CP through UF (box: pre-crack in BL causing crack to cross); d) CR propagating crack leaving PUR-BL along R-direction (box: start of deviation); e-f) IR CP through PVAc (box: adhesive bridging). All scales correspond to 1 mm.

**Table 1:** Number of samples showing crack deflection at BL as a function of adhesive system and angle between load direction and BL (45° - 75°).

| Adhesive | Number of samples with crack deflection at the BL | | |
|---|---|---|---|
| | 45° | 60° | 75° |
| PUR | 11 | 11 | 10 |
| PVAc | 0 | 9 | 12 |
| UF | 3 | 3 | 8 |

As already observed for the 0° samples, UF BLs acted as crack starters, where pre-cracks originated, enhancing the crack propagation through the sample. In cases, where no pre-cracks could be observed, the crack propagation across the BL was so fast that the actual intersection of BL and crack could not be imaged. Only at angles larger than 75° were pre-cracks also detectable in other regions of the BL other than the pure adhesive layer. If the crack was deflected at the BL, then it propagated parallel to it until reaching the next growth ring border or a crack in the BL (Figure 3₁ a-c) were it could cross into the other adherend and further propagate radially through it. Of all studied adhesive systems, UF had the highest MOE contrast compared to spruce, perpendicular to the grain. As a result, shear failure of tracheids was quite common, leading to crack deflection along the BL. To summarize: failure within UF bonded samples was brittle, without a decelerating influence of the BL on the crack propagation. Even though the crack deviated along the BL, it behaved similar to a crack deflected along a growth ring border, exhibiting basically identical microscopic failure mechanisms.

In PUR and PVAc BLs, the crack growth could be slowed down or even stopped by the BL, comparable to the situation at 0°. When deflected, the crack propagated parallel to the BL at least until the next growth ring border or until a defect in the BL was reached (Figure 3$_I$ d-f). In PUR, such defects appeared as stretched pores, and their extensions determined whether the crack crossed the BL into the other adherend. Even a continuation of the deflection parallel to the BL – in the unnotched adherend or a re-crossing of the crack into the notched adherend – could be observed. The crossing of the crack into the unnotched adherend was accompanied by the formation of pre-cracks in both adherends around the BL, stress-whitening in the BL, and the formation of an adhesive bridge. A more precise description of this behavior will be given later. The growth ring borders were preferred zones for crack crossings from one adherend into the other for several reasons: First of all, differences in mechanical properties lead to stress concentrations. Also, residual stresses resulted from differential swelling of EW and LW during the absorption or desorption of water from the adhesive. The different reactions of different BLs towards moisture were recently discussed by Frihart (2009), who showed the ability of adhesives to distribute residual stresses arising from the different swelling behaviors between wood and adhesives. Although focusing on cured BLs, it seems reasonable to assume that these reactions already occur during the bond formation, as the swollen wood was hindered from shrinking back to its original dimensions after the adhesive had solidified. The amount of water introduced depended on the solid content and the curing chemism (see Table , or e.g. Dunky and Niemz 2002). The UF resin and PVAc introduced water into the system, while PUR withdrew some water from the wood for curing. It can be assumed that this led to low residual stresses for PUR and PVAc due to the small dimensional changes for PUR and the low MOE and yield stress of PVAc, when compared to UF, which has a high MOE and additionally causes high moisture induced dimensional changes of the wood substrate.

Rate effects could be observed for PVAc, where cracks propagating at high speeds in the regime above the critical length for stable growth interacted with the BL. In this case, visco-elastic PVAc failed instantaneously. Slow, stable cracks, however, impeded and even stopped at the BL, as there was enough time for plastic stress release and crack tip blunting. The high deformability of the PVAc became visible in an observable relative movement of the two adherends, once the crack entered the BL. Although the final failure pattern suggests that cracks deviated only at angles of 60° or more, the in-situ observations revealed for 45°, cracks started to grow along the BL in combination with the development of stress-whitening and elongation of the adhesive layer. However, with increasing load, the adhesive layer failed and the crack crossed straight into the other adherend without visible deviation in the final failure pattern.

***Load direction and BL at 90°:*** At an angle of 90°, a deflection of the crack along the BL proved most probable. However, it was also observed that cracks stayed within the wood for various reasons. One constellation was found when the R-direction of the adherends was not aligned completely perpendicular to the load direction. Here, the crack could leave its path along the BL and propagate through the wood in the R-direction (Figure 3$_{II}$ d). Another scenario was the crack deviation through an adherend adjacent to the BL. In these cases, the typical failure behavior for SW with unstable crack propagation and pre-cracks in the LW zones could not be observed. Apparently, the adhesive had still an effect at some distance from the actual BL. This zone of influence adds another region to the known bondline composition and its extent, furthermore its dependence on the bondline properties, should be addressed in future investigations.

As cracks take the path of least resistance, they changed from one adherend to the other following the energetically favorable way. Due to differing adhesive properties, the overall crack path along the adhesive interphase was different for the various adhesives.

The behavior of UF bonded samples was similar to that of SW, as pre-cracks mostly appeared in the BL next to LW zones ahead of the main crack (Figures $3_{II}$ a-b). Due to its high stiffness, the UF BL transferred stress directly between both adherends. This means that weak points at some distance from the BL could also form pre-cracks, leading to a failure evolution through SW.

The more flexible PUR and PVAc BLs were also able to deform and dissipate energy. Therefore, pre-cracks were rarely observed distant from the BL. Cracks in wood could be stabilized when they stayed close to the BL, resulting in stable crack propagation even without crossing the BL, as was mentioned before. However, once a crack entered the zone between wood-adhesive interphase and adhesive layer, the majority of cracks followed this interface. It was possible to allocate this failure position due to the formation of adhesive bridges, which seemed to consist mainly of the entire adhesive layer. For further crack propagation, the growth ring borders and their alignment played an important role: (1) When they were alternating (LW zone of one adherend opposite an EW zone of the other), pre-cracks appeared in the BL next to the EW zone, since the low tensile strength of EW led to failure. The crack continued along the BL until reaching the next LW zone, where the pre-cracking and side shifting was repeated (Figure $3_{II}$ e-f). (2) When the growth ring borders of both adherends faced each other, the adhesive was strained, evidenced by stress-whitening next to the LW zones. In absence of relevant BL defects, the samples failed as in SW.

***Additional observations:*** There are several sources for pre-damaging in the BL region. First, the BLs themselves are damaged as a result of their restrained curing (Hass et al. 2011) as described for UF above. Pre-damage may also originate from the bonding process, when rigid LW zones are pressed into soft EW zones (Figure $4_I$ a); EW deformations and even fractures can be observed (Figures $4_I$ c and $4_I$ d). It is possible that uneven sample surfaces prior to bonding (planing) led to pressure peaks, which forced the LW into the EW.

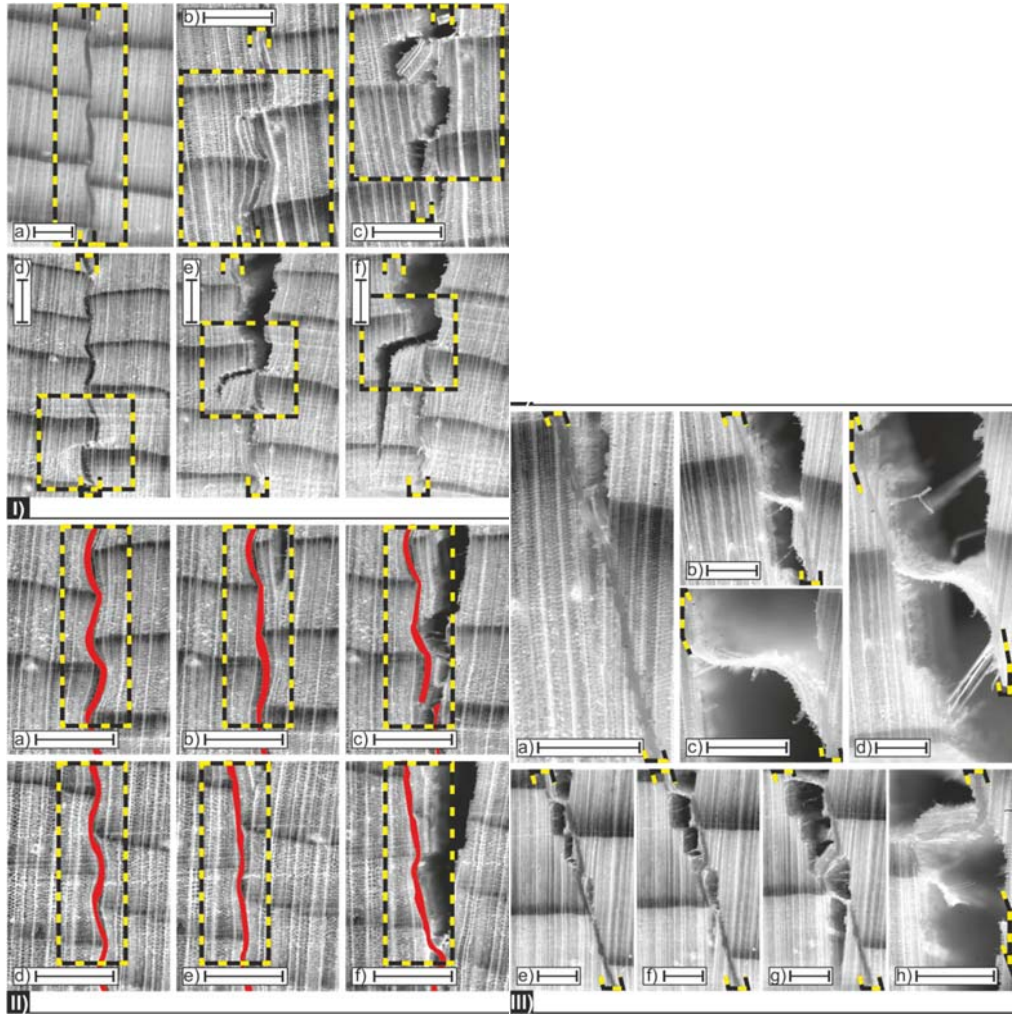

**Figure 4: I)** Different formations of pre-damage: a) Elastic deformations without visible pre-cracks in adherends, no influence on CP; b-c) Pre-damage at LW-EW contact zones acting as pre-cracks for the CP along the BL; d-f) Pre-damage at LW-EW contact zones leading the crack away from the BL. Boxes highlight corresponding fracture zones. **II)** BL deformations in UF (a-c) and PVAc (d-f) at different loading stages. Boxes highlight corresponding positions. **III)** Development of adhesive bridges and fingers in PUR (a-f) and PVAc (g-k). All scales correspond to 1 mm.

Small deformations (Figure 4$_I$ a) did not change the crack path. High deformations or even fractures, however, functioned as pre-cracks, which enhanced the crack propagation (Figure 4$_I$ b-c). They even could direct the crack deeper into the wood, causing the crack to leave the influence zone of the BL and cause instable crack growth through the wood of the adherend (Figure 4$_I$ d-f).

Deformations like elastic compression could even be desirable in case of increased failure strain, because the compression had to be reversed before the tensile stresses arose. For PUR and PVAc, the adhesive layer even remained straight after failure, showing the high amount of plastic deformation of the adhesive layer (Figure 4$_{II}$ d-f). UF BLs however "froze" the wood cells in their compressed state. The stress was transferred directly across the BL and the cells were hindered from relieving the compression. The BL only slightly aligned perpendicular to the load direction during stress and fell back to its compressed position after the crack passed (Figure 4$_{II}$ a-c).

As mentioned previously, the rather flexible PUR and PVAc could also peel off, forming adhesive bridges and fingers. With their high failure strain, they were able to stabilize and slow down the crack propagation (Figure 4$_{III}$). These bridges mainly consisted of the adhesive layer, with additional thin adhesive fingers (PUR see Figure 4$_{III}$ a-d, PVAc see Figure 4$_{III}$ e-h), which connected the two adherends. For UF, the opposite was the case, as wood delaminated from the BL. The only possibility for the formation of stabilizing bridges was given by the wood itself via fiber-bridging.

**Quantitative estimates**

For technical reasons, the tensile strength and the failure displacement were chosen for comparison. Additionally, the displacement was taken as criterion for complete failure until the applied load dropped below 8 N. The tensile strength and failure strain depended primarily on the wood itself, mainly because they were reached before the crack interacted with the BL. Accordingly, values for the different adhesives and load angles were within the range of SW (Figure 5a). The displacement until complete failure confirmed earlier observations, namely, the higher the angle between BL and load direction, the higher is the influence of the adhesive (Figure 5b). For angles < 60°, the differences between the three adhesives were not significant, but a trend was visible with PVAc having the highest failure displacement, followed by PUR and UF. At angles > 75°, the high flexibility of PVAc became more evident and the displacement order was the same as for angles <60°. Although the expected order in flexibility was kept – PVAc, PUR, UF, and SW – the high differences in the adhesives MOE (Table 1) suggested a more pronounced differentiation.

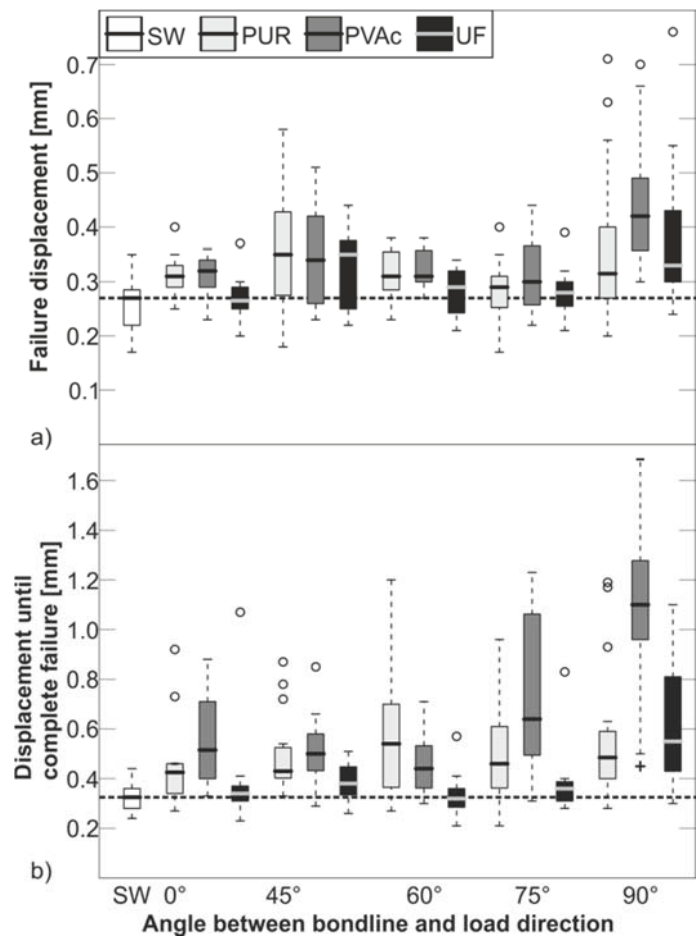

**Figure 5:** Failure displacement (a) and displacement until complete failure (b) for bonded spruce wood and SW under mode I (TR configuration) as a function of adhesive systems and angle between bondline and load direction, compared to SW.

## Conclusions

Depending on the adhesive properties and BL orientation, the crack propagation through a bonded sample can differ considerably from that of SW. Brittle UF BLs provide new crack starters beginning with a curing damage. The crack propagation can be accelerated and shows the same unstable behavior as through SW. BLs of PUR and PVAc can slow down and stabilize the crack propagation compared to SW by forming adhesive bridges between two adherends. They can moderate property differences between tissue types. The adhesive layer itself is able to deform plastically, leading to blunting of the crack tip. Growth ring borders are always preferred positions for a crack to cross between adherends. With suitable adhesives, the failure path and also the duration of bonding until final failure can be increased by stable deflection of the crack along the interphase. The crack propagation is hindered most effectively, when the crack is kept inside the zone of influence of the BL as long as possible. This is in contrast to the traditional belief that the failure of a bonding should occur in the wood part, away from the BL (wood failure). Although this type of failure ensures the integrity of the BL, the positive effects of stable crack propagation along highly dissipative adhesive layers are disregarded, as brittle wood failure is promoted.

The observations in the present paper are a good basis for future failure predictions. In future studies, the effects of pre-compactions on the failure process, as well as the extent of the zone of influence of the BL for different adhesives should be addressed. Additionally, the failure mechanisms of different wood species, including hardwoods, are still waiting for in-depth evaluation.

## Acknowledgements

The authors acknowledge the support under SNF grant 200020_132662 "Micro-Mechanics of Bondline Failure", as well as Dr. Holger Gärtner from the WSL for help with microtome preparations and also thank the companies Purbond and Geistlich Ligamenta for provision of the adhesives.